\documentclass{ws-ijseke}
\usepackage[hyphens, allowmove]{url}
\usepackage{multirow}
\usepackage{booktabs}

\begin{document}

\markboth{NAJAM NAZAR, NORMAN CHEN, and CHUN YONG CHONG.}
{CodeLabeller: A Web-based Code Annotation Tool for Java Design Patterns and Summaries}

%
\catchline{}{}{}{}{}
%

\title{CodeLabeller: A Web-based Code Annotation Tool for Java Design Patterns and Summaries}

\author{NAJAM NAZAR}

\address{Victoria University, Australia\\
najam.nazar@vu.edu.au
}

\author{NORMAN CHEN}

\address{Faculty of Information Technology, Monash University, Australia\\
\email{norman.chen@monash.edu}
}

\author{CHUN YONG CHANG}

\address{School of Information Technology, Monash University, Malaysia\\
\email{chong.chunyong@monash.edu}
}

\maketitle

\begin{history}
\received{(31 December 2022)}
\revised{(12 March 2023)}
\end{history}

\begin{abstract}
While constructing supervised learning models, we require labelled examples to build a corpus and train a machine learning model. However, most studies have built the labelled dataset manually, which in many occasions is a daunting task. To mitigate this problem, we have built an online tool called CodeLabeller. CodeLabeller is a web-based tool that aims to provide an efficient approach to handling the process of labelling source code files for supervised learning methods at scale by improving the data collection process throughout. CodeLabeller is tested by constructing a corpus of over a thousand source files obtained from a large collection of open source Java projects and labelling each Java source file with their respective design patterns and summaries. Twenty five experts in the field of software engineering participated in a usability evaluation of the tool using the standard User Experience Questionnaire online survey. The survey results demonstrate that the tool achieves the (almost) Good standard on hedonic and pragmatic quality standards, is easy to use and meets the needs of the annotating the corpus for supervised classifiers. Apart from assisting researchers in crowdsourcing a labelled dataset, the tool has practical applicability in software engineering education and assists in building expert ratings for software artefacts.
\end{abstract}


\section{Introduction}\label{sec:int}
With the rise of artificial intelligence (AI) in software engineering (SE), researchers have shown how AI can be applied to assist software developers in a wide variety of activities. However, it has not been accompanied by a complementary increase in labelled datasets, which is required in many supervised learning methods. Several studies have been using crowdsourcing platforms to collect labelled training data in recent years such as~\cite{Nazar:2016,Nazar:2022}. However, research such as ~\cite{Nazar:2016,Nazar:2022,Yu:2018} etc has shown that the quality of labelled data is unstable due to participant bias, knowledge variance, and task difficulty.

There are several existing tools that perform design pattern detection (e.g. \cite{Moreno:2012,Nazar:2022}) and code analysis (e.g. \cite{Zhang:2013}). Supervised machine learning for these studies often require the use of a large corpus of source files that have been manually labelled~\cite{Zanoni:2015}. However, labelling design patterns across a large number of source files manually is tedious, as design patterns frequently involve multiple files in their structure, which requires interpreting all involved files in conjunction with each other. Additionally, the need to effectively manage a large number of resulting labels or responses further complicates the problem. For Example, each design pattern usually exhibits fixed characteristics and structure~\cite{Kuchana:2004}. However, human reviewers might not be able to easily reach a consensus on which design patterns are being used in the examined source file~\cite{Nazar:2022}, possibly leading to interlabeller reliability issues~\cite{Nazar:2022}. This could be caused by the source file featuring multiple design patterns, or differing opinions among annotators~\cite{Nazar:2022}. Furthermore, expert contributors are needed to label the source files that are used to train the model or system to identify the design patterns, so that higher Precision and Recall of the classifiers can be obtained while building such systems.

To aid in overcoming these problems and to support research in code quality and static code analysis techniques, we introduce an opensource tool called CodeLabeller. It provides an end-to-end pipeline that streamlines the overall process of manually labelling source files. This streamlining is achieved by leveraging crowdsourcing to obtain a higher number of responses and combining this with efficient response management and automatic collation of responses. CodeLabeller displays one Java source file at a time to each participating contributor and prompts them to identify the design pattern that has been implemented within, to state their confidence level for their selected pattern, as well as to write a brief summary about features of the code in the source file. The tool is evaluated using User Experience Questionnaire - shorter version (UEQ-S) on hedonic and pragmatic standards showing a Good status overall.

\begin{figure*}[htb]%
    \centering
    \includegraphics[scale=0.46]{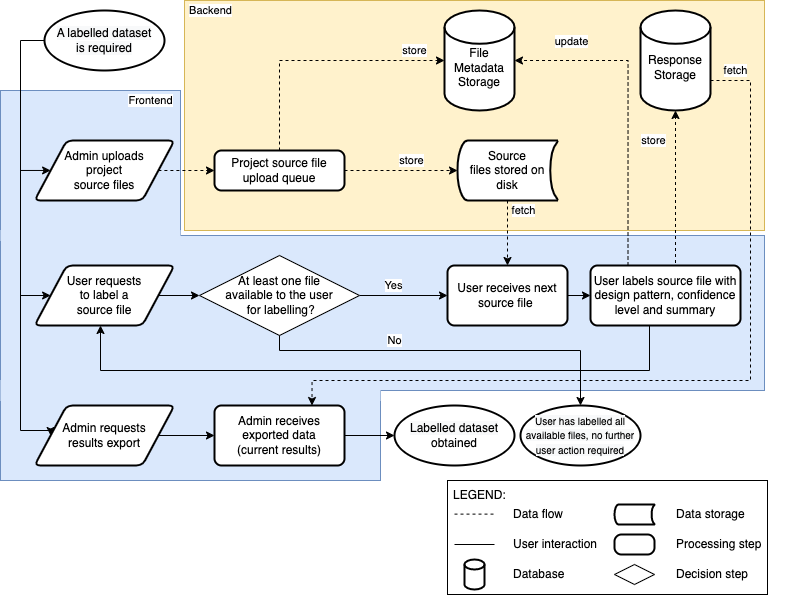}
    \caption{Overall approach and structure of CodeLabeller}\label{fig1}
\end{figure*}

This paper has made the following contributions:

\begin{itemize}
  \item It has provided a tool that future researchers could use to automate the labelling process, especially for supervised learning.
  \item The empirical evaluation demonstrates that the tool has achieved an overall Good rating based on the hedonic and pragmatic quality scales of UEQ-S.
\end{itemize}

The paper is organised as follows: we present the related work and comparison with existing tools in Section~\ref{sec:rw}. Section~\ref{sec:sw} provides a detailed overview of the tool including the framework, structure, working and usage of the tool. We have evaluated our tool in Section~\ref{sec:eval} and impact of the tool in the next Section (i.e.~\ref{sec:impact}). Limitations are discussed in Section~\ref{sec:limitations} and the Section~\ref{sec:conc} concludes our paper.

\section{Related Work}\label{sec:rw}
Design pattern detection and generating automated summaries of source code using supervised learning has been an active area of research in software engineering in recent years. In this section, we discuss some related studies that have utilised supervised machine learning classification for identifying design patterns and generating summaries. We focus on reviewing literature that attempt to build a corpus manually, with the aim to emphasise the need of an automated data collection tool. Unfortunately, many authors do not cite the tool that they used in the annotation process or used a custom (private) tool; thus, we only discuss that are relevant and publicly available.

Zanoni et al.~\cite{Zanoni:2015} exploited the combination of graph matching and machine learning to implement the MARPLE-DPD tool and tested their approach with a manually created corpus called P-MART ~\cite{Zanoni:2015}. Thaller et al.~\cite{Thaller:2019} used the same corpus to build human and machine comprehensible representation based on micro structures (feature maps) to detect design patterns. Recently, Naghdipour et al.~\cite{Naghdipour:2023} discussed the major implications of manually labelling dataset for supervised learning classifiers that are used to identify design patterns. Similarly, for code summarisation, studies such as \cite{Nazar:2016,Sridhara:2010,Moreno:2013,McBurney:2014} applied supervised learning methods to complement manual labelling tasks to generate textual summaries of the target corpus.

There is growing interest in tools that support building annotated corpora for machine learning classification. The BC3: Annotation Software (BC3:AS)\footnote{\url{https://www.cs.ubc.ca/labs/lci/bc3/_annotation/_sw.html}, verified 31/12/2022}, an open-source web annotating tool developed by the Laboratory of Computational Intelligence at the University of British Columbia\footnote{\url{https://www.cs.ubc.ca/labs/lci/bc3/framework.html}, verified 31/12/2022}, is the first software that is extensively used by researchers to annotate learning corpora for email, textual conversations, and bug report summarisations. The tool utilises technologies such as Ruby on Rails and MySQL so that researchers can create a web server that allows them to import and manage an email corpus. This tool has aided studies such as \cite{Rastkar:2014} and \cite{Rastkar:2010} in annotating a corpus for emails, conversations, and bug reports.

As mentioned, BC3:AS is useful for collecting conversation data, such as bug reports and emails. In addition, the studies discussed earlier created tools for design pattern detection and code summaries, but they developed a coprus manually or used a previously available corpus to train and predict the model. There is a need to develop tools that could help to build annotated corpora for source code that could improve the data collection mechanism and improve the quality of the corpus. Nazar et al.~\cite{Nazar:2022} observed that building a corpus using a tool improves the quality and collection of the dataset compared to manually created corpus as they reported an increased Precision and Recall score compared to studies that have manually created corpora, in their study.

Our Codelabeller is a publicly available code annotation tool that can be easily customisable, and can facilitate future researchers in building annotated corpora of source code for supervised learning. Our Codelabeller tool is inspired by BC3:AS, and it differs from BC3:AS in many ways, as it is robust and solely for annotating source files. It differs with BC3:AS in many ways such as, it has an admin panel that helps in populating the corpus with any programming languages - though we have only tested it with Java programming language. Other related features are discussed in Section~\ref{sec:sw}. We have described the difference with the tool in Table~\ref{tab:diff}.

\begin{table*}[]\centering
\caption{Differences between BC3 and CodeLabeller}
\label{tab:diff}
\resizebox{\textwidth}{!}{%
\begin{tabular}{ll}
\toprule
\textbf{BC3}                     & \textbf{CodeLabeller}\\
\midrule
Annotate Emails and conversation & Annotates source code                                                                                \\
Developed in Ruby on Rails       & Developed in Node JS                                                                             \\
Predefined email threads         & Customised dataset in any programming language    \\
No Confidence labelling          & Confidence labelling.\\
\bottomrule
\end{tabular}
}
\end{table*}

\section{Tool Description}\label{sec:sw}
In this section, we present a detailed overview of the tool, discussing tool implementation, usage, features, and the labelling process with examples. The Figure~\ref{fig1} provides an overview of our codelabeller tool.

\subsection{Implementation}\label{subsec:impl}
CodeLabeller is a web-based application consisting of a front-end browser client written using the Angular framework and a back-end server written in Node.js which stores application data in a MySQL database. The implementation code is available at \url{https://github.com/najamnazar/codelabeller}, whereas the tool is accessible at \url{https://codelabeller.org}.

\subsection{Tool Usage}\label{subsec:tool}
When starting a new labelling initiative, researchers are able to upload batches of source files on a project-by-project basis. Once these source files have been successfully uploaded, CodeLabeller automatically manages the corpus. During the labelling process, instead of assigning source files to users in either a random or fixed order, CodeLabeller prioritises assigning contributors to source files that currently have the smallest difference between the number of responses required and the number of responses currently received. This allows for more source files to be fully labelled much earlier, thus avoiding the problems associated with the random labelling of larger source file corpora, where it might take a longer time before any source file receives enough labels. At any time throughout the labelling process, the source files can also be individually deactivated and set to stop accepting further responses. Contributors are presented with one Java source file at a time and will be requested to indicate which design pattern (or lack thereof) is featured in the code, their confidence level, and to write a brief summary about the code in the current file. The contributor will choose from a list of design patterns specified by the researchers who use CodeLabeller. When a contributor has identified a Java source file as containing a design pattern that has not been included in the list specified by the researcher, the contributor may manually type in the name of the unlisted design pattern. The contributor may additionally add other notes about the presented source file.

\begin{figure}[htb]%
    \centering
    \includegraphics[scale=0.25]{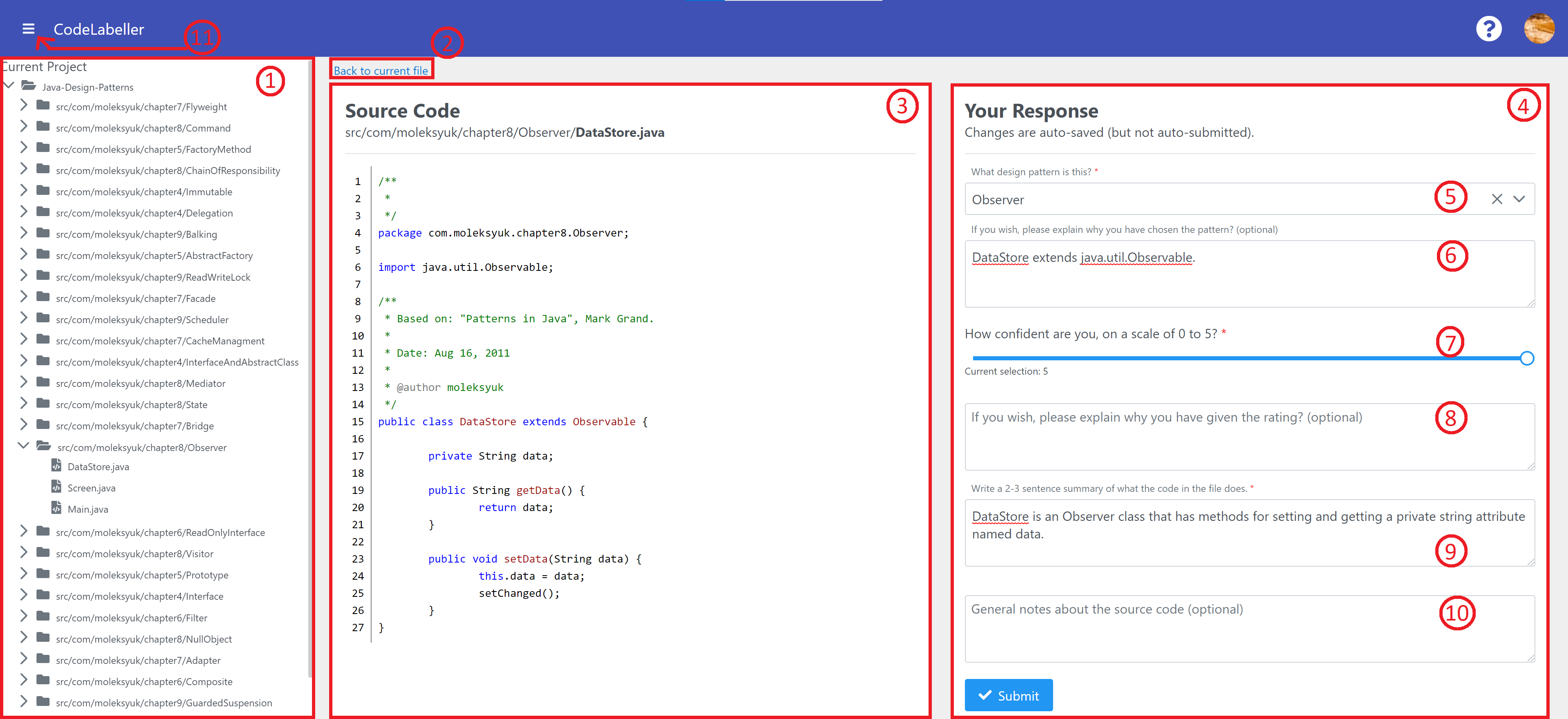}
    \caption{The labelling dashboard of CodeLabeller, showing project navigation, source code viewer, and response input panes}		\label{fig2}
\end{figure}

In all cases, the contributors must specify how confident they are with the label and the explanations given for each response. This is important to address a common issue in crowdsourcing where contributors may come from different backgrounds and proficiency levels, and quality control of responses need to be ensured. The confidence level parameter attached to each response will allow researchers to easily identify contributors that are considered to be highly proficient (expert contributors) and to invite these individuals to participate in more labelling initiatives in the future. In addition, this enables machine learning classifiers to give higher weighting to the responses made by these expert contributors. The combined efforts lead to the construction of higher quality labelled datasets being produced at the end of a labelling initiative.

Some source files may contain more than one design pattern. In this instance, the contributor will label the presented source file with the design pattern that they feel is (or possibly) most relevant. Additionally, the implementation of some design patterns (such as the Observer pattern) may involve more than one source file. If a presented source file forms part of an overall design pattern, contributors will label that file with the design pattern that it is part of. In this situation, CodeLabeller assists contributors by enabling them to navigate to other source files within the same project and view their respective code.

Upon submission of a response, CodeLabeller will then present the contributor with another source file, and the process repeats until the contributor has labelled all available source files. At any time, contributors can view all submitted responses and can go back to any previous response to review and update wherever necessary (see Figure \ref{fig4}). All response data are presented in tabular form and can be sorted, filtered, and downloaded. CodeLabeller allows administrators to download CSV responses containing only responses that match their search/filter parameters. We discuss the whole process with examples in subsequent sections.

\subsubsection{Starting a lebelling initiative}
When a labelled dataset is required by researchers, an instance of CodeLabeller is deployed and the dataset corpus is uploaded as a .zip archive file to CodeLabeller by the researchers (who function as CodeLabeller administrators).

A corpus consists of several directories, where each directory represents a single Java code repository/project. A project directory may contain further subdirectories within it, with any number of Java source files within each project. When CodeLabeller receives an uploaded .zip archive file, each file within the .zip file will be queued for processing. Only files that are Java source files will be added to the corpus, other files will be ignored and will not be added to the corpus. This could be customised to other programming languages such as Python.

\subsubsection{Management of the Corpus}
Each project can be set to active or inactive, which determines if the Java source files contained within it can be assigned to users for labelling. Each Java source file can be set to active or inactive, which determines visibility of the file in the project navigation pane during the labelling process, and whether or not to accept new responses. Additionally, the target number of responses each Java source file requires can be individually set. These parameters can be changed at any time throughout the labelling initiative via the administration dashboard.

If additional Java source files are required to be added to the corpus halfway through the labelling initiative, these can be compressed into another .zip archive file and uploaded to CodeLabeller at any time to be processed and added to the corpus as described in (1).
\begin{table}[]
\centering
\caption{Labelling of the corpus}
\label{tab:lab}
\resizebox{\textwidth}{!}{%
\begin{tabular}{llll}
\toprule
\textbf{Java Source File} & \textbf{Responses Obtained} & \textbf{Target \# of Responses} & \textbf{\# Responses Still Needed} \\
\midrule
A & 3 & 5 & 2 \\
B & 0 & 3 & 3 \\
C & 1 & 3 & 2 \\
D & 2 & 3 & 1 \\
E & 2 & 5 & 3 \\
\bottomrule
\end{tabular}%
}
\end{table}
\subsubsection{Lebelling of the Corpus}
CodeLabeller will assign each user one Java source file at a time from the set of files assigned to the labeller. In order for a Java source file to be assignable to a user, both the file and its associated project must be set to active and accepting responses.

Users label their currently assigned Java source file and while doing so, may freely browse other active Java source files within the current Java source file's project.

CodeLabeller assigns Java source files to users in a manner that prioritises files that need the least number of responses to reach its target number of responses. For example, given files A-E as follows, and two users A and B (who have not yet labelled any of the files A-E) as shown in Table~\ref{fig2}:

File D will be assigned to the next available user (say, user A) as it is currently the only file that has the least number of responses still required to achieve its response target. Once file D is labelled by user A as shown in Table~\ref{tab:lab2}:

\begin{table}[]
\centering
\caption{Labelling of the corpus - part 2}
\label{tab:lab2}
\resizebox{\textwidth}{!}{%
\begin{tabular}{llll}
\toprule
\textbf{Java Source File} & \textbf{Responses Obtained} & \textbf{Target \# of Responses} & \textbf{\# Responses Still Needed} \\
\midrule
A & 3 & 5 & 2 \\
B & 0 & 3 & 3 \\
C & 1 & 3 & 2 \\
D & 0 & 3 & 0 \\
E & 2 & 5 & 3 \\
\bottomrule
\end{tabular}%
}
\end{table}

Either file A or file C will now be assigned to the next available user (say, user B), as these two files are now both equally nearest to completion (2 more responses to completion).  Assuming that file C is assigned to user B and a response is then submitted, file C will now be the file closest to completion (1 more response required) as described in Table~\ref{tab:lab3}.

When user B requests another file to label, CodeLabeller assigns file A to user B instead of file C even though file A requires 2 more responses to completion compared to file C that requires 1 more response. File C has already been labelled by user B and hence file A now becomes the file nearest to completion for user B. The next user to request for a file to label that has not labelled file C will then be assigned file C (e.g. user A). 

A particular user will continue to be assigned source files to label, until all files have been labelled by that user, and this continues until all users have labelled all files or all files have achieved their respective response targets.

\begin{table}[]
\centering
\caption{Labelling of the corpus - part 3}
\label{tab:lab3}
\resizebox{\textwidth}{!}{%
\begin{tabular}{llll}
\toprule
\textbf{Java Source File} & \textbf{Responses Obtained} & \textbf{Target \# of Responses} & \textbf{\# Responses Still Needed} \\
\midrule
A & 3 & 5 & 2 \\
B & 0 & 3 & 3 \\
C & 2 & 3 & 1 \\
D & 0 & 3 & 0 \\
E & 2 & 5 & 3 \\
\bottomrule
\end{tabular}%
}
\end{table}
\subsubsection{Viewing and exporting responses}
All responses can be viewed and exported by a CodeLabeller administrator at any time throughout the labelling initiative.

\subsection{Notable Features}\label{subsec:nf}
Some notable features of CodeLabeller include minimising the loss of unsubmitted data by automatically saving draft responses locally in the browser each time a change is made, as well as storing all past versions (edit histories) of each submitted response, which only administrators can view. CodeLabeller also features a RESTful API and utilises OAuth 2.0 as a user authorisation mechanism, which aids interoperability with other systems as CodeLabeller is further improved in the future. We plan to extend the authentication and authorisation with other providers such as Microsoft in future releases of CodeLabeller.

\subsection{Labelling Dashboard}\label{subsec:dashboard}
Figure \ref{fig2} shows CodeLabeller’s main page, which consists of three major UI components. Contributors who have successfully logged in will be brought to the dashboard.

Project navigation pane (1). As some design patterns may span over several source files, contributors may freely navigate and view any other source file within the same project where their currently assigned source file is located. When browsing another source file, contributors may also quickly return to their currently assigned source file (2). If required, the contributor may also temporarily hide this pane by clicking on the left-menu button on the toolbar (11).

Source code viewer pane (3). The source code of the file that the contributor is currently browsing will be shown here. Contributors will see the full file path of the currently viewed source file, and the code displayed in this pane is syntax-highlighted. For lengthy source files containing many lines of code, the contributor may scroll through the code vertically and horizontally.

Response input pane (4). This is the main response area where contributors submit a new response for a source file that they have not labelled before or update an existing response for a previously labelled one. As mentioned in Section 3.2, any draft responses for the source file currently being labelled will automatically be saved locally on the contributor’s web browser to avoid loss of work. Here, contributors select a design pattern (or a lack thereof) (5), explain why they have selected that option (6), provide their confidence level (7), explain why they have selected that confidence level (8), write a short summary of the features in the presented code ( 9 ), and provide any general notes for it (10).
\subsection{Administration Dashboard}\label{subsec: ad_dashboard}
\begin{figure*}[htb]%
\centering
\includegraphics[scale=0.25]{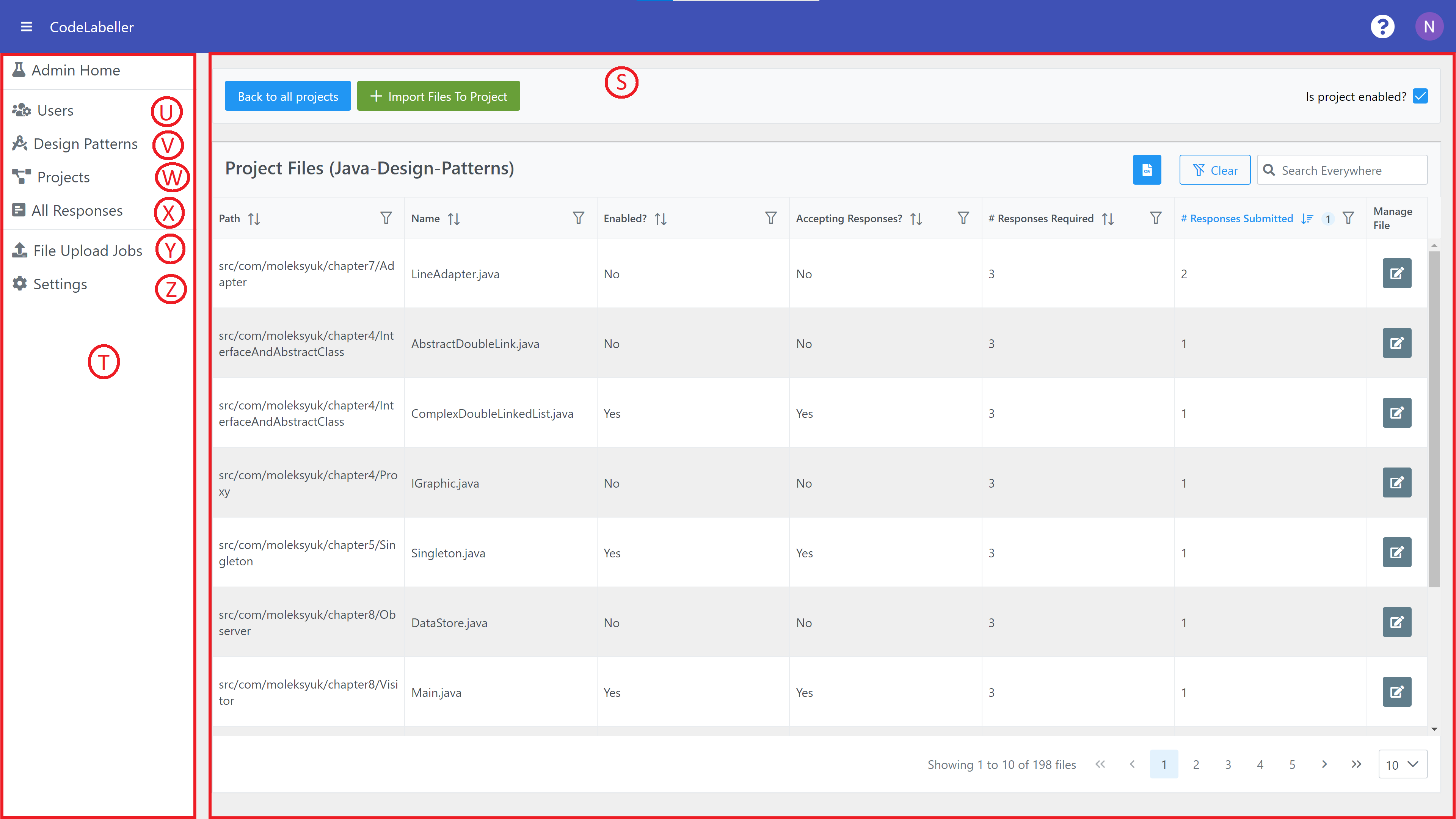}
\caption{The administration dashboard for administrators, showing the file management functionality.}\label{fig3}
\end{figure*}

CodeLabeller also features a dashboard for administrators, as shown in Figure \ref{fig3}, which allows the administration of various aspects of the labelling process.

Administration functionality area (S). Once an administrator has selected an administrative functionality, the relevant user interface and controls for the selected functionality will appear here.

Administration navigation pane (T). Available administrative tasks include contributor/user management (U), design pattern management (V), project and source file upload/management (W), viewing/downloading all responses submitted by all contributors (X , and illustrated in Figure \ref{fig4}), viewing the status of project file uploads (Y), and managing other settings of CodeLabeller (Z).

\subsection{''My Responses'' Page}
\begin{figure*}[htb]%
\centering
\includegraphics[scale=0.25]{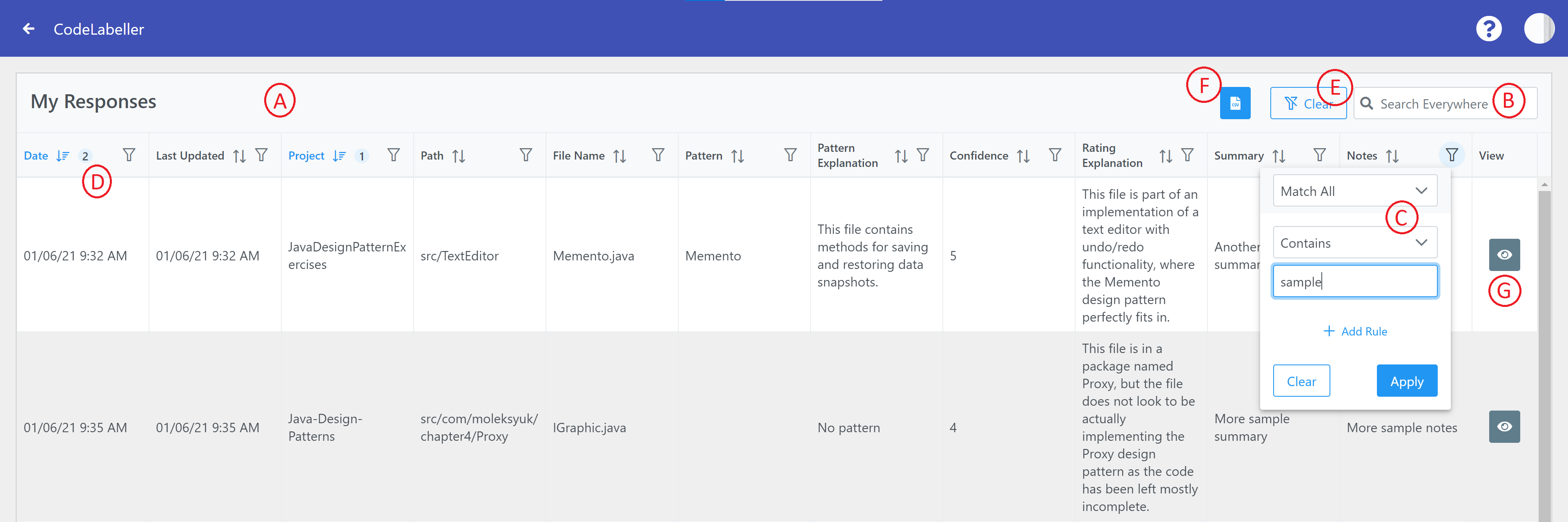}
\caption{The "My Responses" page for a contributor, showing previously submitted responses and other controls.}\label{fig4}
\end{figure*}

In Figure \ref{fig4}, contributors can view a table of all their responses (A). The latest version of the response for each source file will be shown. A global search can be performed on all columns in the responses table (B), and each column can also be individually filtered (C). The responses can be sorted by one or more columns (D). At any time, any applied filter, sorting, and search criteria can be cleared (E) and the current state of the response table can be downloaded by contributors as a CSV file (F). Contributors may also navigate back to the labelling dashboard (Figure \ref{fig2}) to view/update any of their previously submitted responses (G).

\section{Evaluation of the Tool}\label{sec:eval}
We have performed human evaluation through a usability study for evaluating the efficacy of the tool. In the subsequent sections, we discuss each criterion one by one.

\subsection{Participants and Procedure}\label{subsec:part}
In total, 25 participants participated in the user study through an online survey. The study obtained ethics approval from the university before it started.

The evaluation was carried out using Likert scales from 1 to 7, with open-ended questions to understand user preferences in the design or practices to improve the overall usability. The feedback received will be used to refine the tool to make it more user-friendly in future releases.

The evaluation survey was delivered in the form of a custom-built website that provides a step-by-step guide on the evaluation process and the anonymous online survey. It requires around 10 - 15 minutes to complete, including the labelling of the files. The participants need to use the tool, label some design pattern files provided in the tool link, write short summaries and identify what pattern does a file contain. We expected that the participants should label at least 10 files to better evaluate the tool. However, this number varies as some participants prefer to  label more than 10 files whereas one participant labelled only 5 files.

\subsection{Study Instrument}
The study survey is adapted from the User Experience Questionnaire (UEQ-S)\footnote{\url{https://www.ueq-online.org/}. UEQ-S is a shorter version of the original UEQ survey that uses 26 items.}\cite{Schrepp:2017}. The UEQ-S is a widely used questionnaire to measure the subjective impression of users towards the user experience of products \cite{Schrepp:2017}. The UEQ-S includes 8 items on a 7-level Likert scale that are grouped into six categories. The UEQ-S instrument highlights user experience (attractiveness) based on two dimensions, the pragmatic dimension that assesses perspicuity, efficiency and dependability of the tool, and the hedonic dimension that focuses on the novelty and stimulation of the tool.

\subsection{Results}
A total of 25 participants consented to participate in the study. The participants aged 26 to 48 
, the majority were male 
and held a tertiary level qualification – Bachelor, Masters and Doctorate degrees as shown in Table~\ref{tab1}. The majority of the participants possess over 5 years of Java programming experience and familiarity with software design patterns except two participants who have little to one year experience in design patterns. Most of the participants labelled at least 10 files through the tool.

\begin{table}[htb]
\centering
\caption{Demographic characteristics of the survey respondents n = 20}
\label{tab1}
\begin{tabular}{@{}lll@{}}
\toprule
\textbf{Characteristics}                    & \textbf{Participants Groups} & \textbf{\# of Participants} \\ \midrule
\multirow{3}{*}{Age}                        & 20-30 years                  & 6                           \\
                                            & 30-40                        & 10                           \\
                                            & 40-50                        & 9                           \\ \midrule
\multirow{2}{*}{Gender}                     & Male                         & 15                           \\
                                            & Female                       & 10                           \\ \midrule
\multirow{3}{*}{Education}                  & Bachelor's Degree            & 4                           \\
                                            & Master’s Degree              & 10                           \\
                                            & Doctoral Degree              & 11                           \\ \midrule
\multirow{3}{*}{Java Experience}            & 1 - 5 years                  & 7                           \\
                                            & 5 - 10 years                 & 6                           \\
                                            & 10+ years                    & 12                           \\ \midrule
\multirow{3}{*}{Design Patterns Experience} & 1 - 5 years                  & 12                           \\
                                            & 5 -10 years                  & 10                           \\
                                            & 10+ years                    & 3                           \\ \midrule
\multirow{3}{*}{\# of labelled files}       & 1 - 10 files                 & 8                           \\
                                            & 10 - 20 files                & 15                           \\
                                            & 20 + files                   & 2                           \\
\bottomrule
\end{tabular}
\end{table}

\begin{figure*}[htb]%
\centering
\includegraphics[scale=0.34]{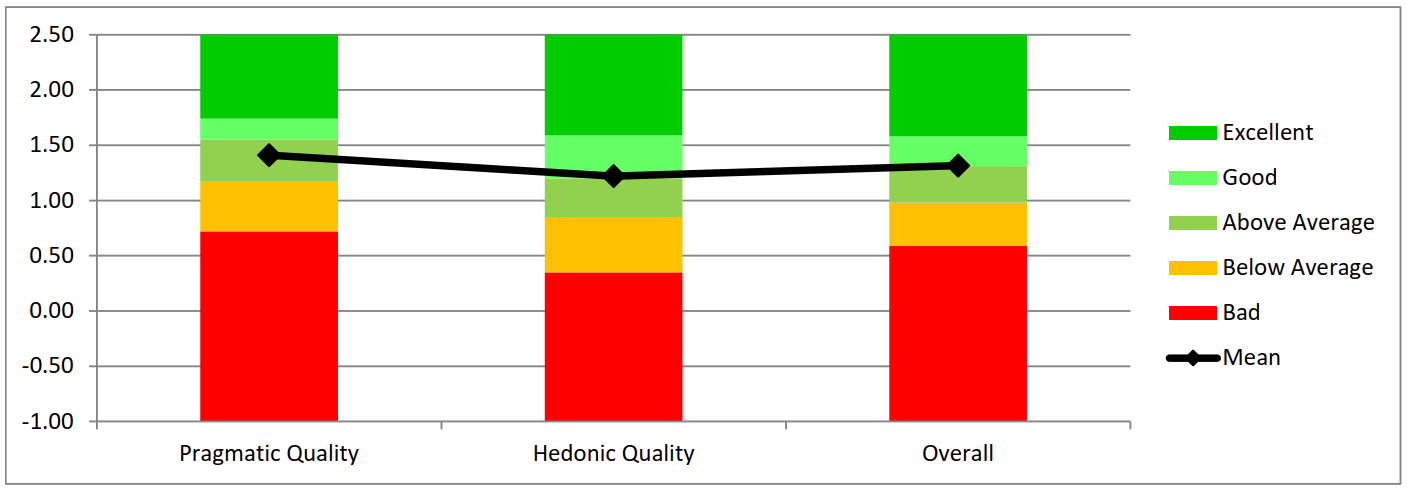}
\caption{The overall quality of the tool based on pragmatic and hedonic qualities of the UEQ-S}\label{fig6}
\end{figure*}

Overall participants demonstrated that the tool is Good to use except one participant who raised concerns about every aspect of UEQ-S for the tool -  this could possibly be that the participant has novice level experience in software design pattern and found it difficult to make a judgement about what pattern is used in the java file. Additionally, the participant labelled less than 5 files, which was less than the minimum threshold of labelling at least 10 files for better judgement for the tool. Usually, the participants having little or no experience (less than a year) with design patterns rated the efficiency and supportive aspects of the survey low. This low score by novice experts has affected the pragmatic and hedonic quality of the tool reducing the overall tool quality from Good to just at the edge of the good in some aspects of hedonic and pragmatic quality aspects as shown in table~\ref{tab2}. The Figure~\ref{fig6} further demonstrates it in the graphical form.

\begin{table}[!ht]
    \centering
    \caption{UEQ-S results}
    \label{tab2}
    \begin{tabular}{ll}
    \toprule
    \textbf{Short UEQ Scales} & \textbf{Mean}  \\
    \midrule
        Pragmatic Quality & 1.41  \\ 
        Hedonic Quality & 1.22  \\ \hline
        Overall & 1.32 \\
    \bottomrule
    \end{tabular}
\end{table}

\section{Impact}\label{sec:impact}
Apart from creating a labelled dataset for supervised learning methods, there are several scenarios that could benefit from the use of CodeLabeller such as for software engineering education, where it can serve as a practical platform. As an example, for students who are learning about design patterns, educators can upload source files containing known design patterns and request students to label these files as part of their formative and/or summative assessments. Educators will then be able to easily collect and collate student responses, thus facilitating a deeper insight into students’ level of understanding. The overall approach used by CodeLabeller can also be extended to bring benefits to other software engineering domains such as code reviews and effort estimation that require the recruitment of expert groups to manually label various software artefacts such as bug reports, and the subsequent aggregation of these expert opinions.
\section{Limitations}\label{sec:limitations}
The study has some limitations. First, it was difficult to hire design pattern expert participants for conducting a through survey. We managed to find ten participants to fulfil the minimum threshold value of 10 participants; however, some of them possess very little understanding of design patterns. Second, the inability of the researcher to physically conduct a workshop to monitor user interactions and receive real-time feedback. We have to rely on the online survey and in most cases the participants provided no comments or suggestions in regards to the tool. Third, it was difficult for us to figure out if the participants understood completely the purpose of the tool. Although we have provided them with a help menu, we still noticed that some participants took extra time to understand the tool. Finally, due to the short duration of the project, the participants were provided only a week’s time to conduct a comprehensive evaluation. However, since this is a pilot evaluation, we aim to perform a more comprehensive future evaluation with practical interactions to determine the tool’s effectiveness in supporting the labelling data for supervised learning based studies. We plan to counter aforementioned limitations in the future studies.

\section{Conclusion and Future Work}\label{sec:conc}
In this paper, we present CodeLabeller, a web-based tool that allows annotations for Java source files to be obtained on a crowdsourced basis. These annotations can then be used in supervised machine learning efforts, for instance, automatically detecting design patterns in source code. Apart from using the tool to create a labelled dataset, it can also be used by educators to teach machine learning classification and software design in a classroom setting. In addition, researchers can use the tool to gather expert opinions and build estimations for software artefacts. The usability study based on the UEQ-S model determines that the tool is useful and achieved an overall Good value on the quality scales of excellent to bad. 

For the next release of CodeLabeller, a few enhancements and extensions are being considered that are adding a section where contributors can add method summaries, navigate through code definitions by clicking on the file name and ability to add other software artefacts. Furthermore, we plan to convert the tool to an eclipse plugin to target wider audience and usage. We are also considering to add an admin panel where users can heuristically generate the required information for the supervised experiments they aim to conduct such as inter rating agreements.

\bibliographystyle{ws-ijseke}
\bibliography{ws-ijseke}

\end{document}